\documentclass[a4paper,12pt,english]{article}
\usepackage{amsfonts}
\usepackage{graphicx}
\usepackage{amsmath}
\usepackage{color}

\newcommand{\be}{\begin{equation}}
\newcommand{\ee}{\end{equation}}
\begin{document}


\begin{titlepage}
\begin{center}

\noindent{{\LARGE{A closer glance at black hole pair creation}}}

\smallskip
\smallskip

\smallskip
\smallskip

\smallskip
\smallskip
\smallskip
\smallskip
\noindent{\large{Andr\'es Anabal\'on$^1$, Sasha Brenner$^2$, Gaston Giribet$^2$, Luciano Montecchio$^2$}}

\smallskip
\smallskip

\smallskip
\smallskip

\centerline{$^1$ Universidad Adolfo Ib\'a\~nez, Dep. de Ciencias, Fac. de Artes Liberales} 
\centerline{{\it Av. Padre Hurtado 750, Vi\~na del Mar, Chile.}}
\centerline{$^2$ Politecnico di Torino, Dipartimento DISAT. Corso Duca degli Abruzzi 24, 10129 Torino, Italy.}
\smallskip
\smallskip
\centerline{$^2$ Physics Department, University of Buenos Aires FCEyN-UBA and IFIBA-CONICET}
\centerline{{\it Ciudad Universitaria, pabell\'on 1, 1428, Buenos Aires, Argentina.}}

\end{center}

\bigskip

\bigskip

\bigskip

\bigskip

\begin{abstract}
We consider accelerated black hole horizons with and without defects. These horizons appear in the $C$-metric solution to Einstein equations and in its generalization to the case where external fields are present. These solutions realize a variety of physical processes, from the decay of a cosmic string by a black hole pair nucleation to the creation of a black hole pair by an external electromagnetic field. Here, we show that such geometries exhibit an infinite set of symmetries in their near horizon region, generalizing in this way previous results for smooth isolated horizons. By considering the limit close to both the black hole and the acceleration horizons, we show that a sensible set of asymptotic boundary conditions gets preserved by supertranslation and superrotation transformations. By acting on the geometry with such transformations, we derive the superrotated, supertranslated version of the $C$-metric and compute the associated conserved charges.We also consider other physical scenarios, including accelerated black holes in AdS and binary black hole systems.
\end{abstract}
\end{titlepage}


\section{Introduction}

A pair of black holes with opposite, constant proper accelerations is described by the so-called $C$-metric \cite{C0, Podolsky}, a gravitational instanton solution to the vacuum Einstein equations \cite{Weyl, Solutions}. This solution describes a pair of two accelerated black holes pulled apart by the action of two semi-infinite cosmic strings, each of them attached to one black hole and extending all the way to infinity. This type of configuration has very interesting applications: For example, this can be used to describe the process through which a cosmic string decays by nucleation of two black holes. Also, it can be used to study processes that represent self-gravitating analogs of the Schwinger effect: In fact, both magnetically and electrically charged generalizations of the $C$-metric in presence of external fields are known \cite{Ernst}, these being given by axisymmetric electro-vacuum solutions to the Einstein-Maxwell equations. Then, it is possible to resort to the Euclidean path integral method and compute the production rate of charged black holes in an external electromagnetic field \cite{Melvin}, a kind of process that has been extensively studied in the literature \cite{Grifits, Grifits2, Hawking, Astorino}, especially in relation to the microscopic description of black hole thermodynamics \cite{Grifits2, Hawking}.

The thermodynamics of $C$-metric type solutions has recently been revisited from a modern perspective \cite{C2, C4, C11, C12}, and asymptotically, locally AdS versions of these geometries \cite{Holo0} have been studied in the context of holography \cite{Holo1, Holo2, Holo3}; see also references therein and thereof\footnote{See also references \cite{unotro1, unotro2, unotro3, unotro4, unotro45, unotro5, unotro6, unotro7, unotro8}, where the $C$-metric and related geometries are studied in different contexts}. In a recent paper \cite{Strominger}, Strominger and Zhiboedov consider the $C$-metric as a working example to investigate the action of the Bondi-Metzner-Sachs (BMS) superrotation symmetry \cite{Glenn, Glenn2} that appears in the asymptotically flat spacetimes at null infinity. Superrotations act on the celestial sphere as local conformal transformations, and this introduces singularities whose physical meaning remained unclear. By analyzing the process of black holes nucleation and cosmic string decay, Strominger and Zhiboedov showed that the early and late time regimes of such a process lead to two distinct vacua that differ from each other by a finite superrotation, showing that the latter transformations are necessary to describe physical mechanisms of this sort. 

Here, taking the result of \cite{Strominger} as a motivation, we will study superrotations and supertranslations symmetries of the $C$-metric but from the near horizon perspective. As shown in \cite{DGGP1}, the near horizon regions of many gravity solutions also exhibit a BMS-type asymptotic symmetry that includes superrotations. We will investigate how this symmetry acts in the case of the $C$-metric, both at the black hole horizon and at the acceleration horizon. We will show that the $C$-metric solution can actually be accommodated in the set of near horizon boundary conditions that are preserved by an infinite-dimensional symmetry algebra. The zero-modes of the Noether charges associated to such infinite symmetry correctly reproduce the Wald entropy of the $C$-metric horizons. Then, we will construct the supertranslated and superrotated versions of the $C$-metric explicitly.

The paper is organized as follows: In section 2, we will review the main geometrical features of the $C$-metric solution as well as its physical interpretation. In section 3, we will consider the near horizon symmetries of the $C$-metric and the corresponding charge algebra. In section 4, we will construct both superrotated and supertranslated versions of the $C$-metric solution. In section 5, we discuss the case of accelerated black holes in AdS spacetime, which echibits features qualitatively different from those of the asymptotically flat case. In section 6, with the intention of showing how the near horizon analysis can be extended to gather physical scenarios of different sorts, we address a case that describes a black hole binary system in equilibrium. For this case as well, we will find results compatible with the black hole thermodynamics.

\section{The $C$-metric and black hole pairs}

In a suitable system of coordinates, the $C$-metric takes the following form \cite{Teo}
\begin{equation}\label{La1}
ds^2= \frac{1}{\alpha^2 (x+y)^2} \left( -f(y)\, d\tau ^2 + \frac{dy^2}{f(y)} + \frac{dx^2}{g(x)} + g(x)d\phi^2 \right)\\
\end{equation}
with
\begin{equation}\label{La2}
     f(y) = -(1-y^2)(1-2\alpha M y)\, , \ \ \  g(x) = (1-x^2)(1+2\alpha M x)\, ,
\end{equation}
where $\tau \in \mathbb{R}$, $x\in \mathbb{R}$, $y\in \mathbb{R}$, and where $\phi \in [-C\pi,C\pi]$ is a periodic coordinate. The value of $C$ controls the angular deficit of a conical singularity that the geometry exhibits at $x^2=1$. This codimension-2 defect is ultimately interpreted as two semi-infinite cosmic strings that pull the black holes and take them apart, accelerating each of them in opposite directions. Each string is attached to one black hole and it extends all the way to infinity, pinching off the null infinity at antipodal points of the celestial sphere. $C$ thus measures the tension of the strings. In (\ref{La1}), $\alpha$ and $M$ are two arbitrary parameters that are ultimately associated to the acceleration and the mass of the solution, respectively. The metric functions in (\ref{La2}) satisfy $g(x)=-f(-x)$, and the metric is singular at the {conformal infinity} $x+y=0$, where the conformal factor diverges. 

In order to make the interpretation of the parameters $\alpha$ and $M$ clear, it is convenient to consider the change of coordinates
\begin{equation}\label{coor}
    x = \cos \theta \ , \ \ \ y = \frac{1}{\alpha r}\ , \ \ \ \tau = \alpha t
\end{equation}
which is well defined provided one restricts the analysis to the region $ -1 \leq x \leq 1$, $y>0$. In fact, coordinates ($t, r, \phi , \theta $) cover only part of the manifold, where one sees only one black hole. In this coordinates, the metric reads
\begin{equation}\label{C-metric}
    ds^2 = \frac{1}{(1+\alpha r \cos \theta )^2} \left( -F(r)\, dt^2 + \frac{dr^2}{F(r)} + \frac{r^2 d\theta^2}{G(\theta)} + G(\theta)r^2 \sin^2\theta \, d\phi^2 \right)
\end{equation}
with
\begin{equation}
F(r) = (1 - \alpha^2 r^2) \Big( 1- \frac{2M}{r}\Big) , \quad \quad G(\theta) = 1 + 2\alpha M \cos\theta \, .
\end{equation}
Now, it becomes evident that the case $\alpha=0$ corresponds to Schwarzschild solution with mass $M$ (here, we use natural units $G=c=1$). On the other hand, the case $M=0$ with $\alpha $ arbitrary gives a solution that is diffeomorphic to Minkowski; more precisely, it reduces to Rindler spacetime as perceived by an observer at $r=0$ with proper constant acceleration $\alpha $. The metric with $M\neq 0$ is singular at $r=0$, and it exhibits two Killing horizons at $r_+ = 2M$ and $r_a = 1/\alpha$; these correspond to the event horizon and the acceleration horizon, respectively. In order to guarantee the spatial character of the angular coordinates $\theta$ and $\phi$, it is necessary to impose $2M < 1/\alpha$, in such a way that $G(\theta) > 0 $ for all values of $ \theta$. The latter condition is equivalent to demanding the two horizons to be disjoint.

For the solution not to exhibit angular deficit at the poles of the constant-$r$, constant-$t$ surfaces, it is necessary to fix the value of $C$ to ensure the right periodicity in $\phi $, either at $\theta =0$ or at $\theta =\pi $. Since for $M\neq 0$ both poles have different conicity, it is only possible to cure only one conical singularity and not both. For instance, we could take $C=C_{\theta =0} \equiv (1 +2\alpha M)^{-1}$ which suffices to make the surface smooth at $\theta = 0$ at the price of fixing the angular deficit at $\theta=\pi $ to be $\delta_{\theta=\pi}= \frac{8\pi \alpha M}{1+2\alpha M}$. This is precisely the codimension-2 defect we mentioned above: The angular deficit extends from the event horizon all the way to infinity, describing the cosmic string that pulls the black hole. In this scenario, it is the string what provides the black hole with its proper acceleration. However, we could also think of other mechanism: For example, we could consider the system in presence of a uniform external magnetic field, and then think of a pair of magnetically charged black holes being spontaneously created by tunneling, as if it was a self-gravitating analog of the Schwinger effect. This amounts to consider the magnetically charged version of the $C$-metric in the background of external magnetic field $B$, which is given by \cite{Ernst}
\begin{equation}\label{La8}
    ds^2= \frac{\lambda^2(x,y)}{\alpha^2 (x+y)^2} \left( -f(y) \, d\tau ^2 + \frac{dy^2}{f(y)} + \frac{dx^2}{g(x)} \right) + \frac{g(x)}{ \lambda^2(x,y)  \alpha^2 (x+y)^2}d\phi^2 
\end{equation}
with
\begin{equation}\lambda(x,y) = \left( 1 - \frac{Bp}{2} x \right)^2 + \frac{B^2}{4\alpha^2 (x+y)^2} \hspace{0.3em} g(x),
\end{equation}
and with the new functions
\begin{equation} 
f(y) = -(1-y^2)(1-\alpha r_+ y)(1-\alpha r_- y), \quad \quad g(x) = (1-x^2)(1+\alpha r_+ x) (1+\alpha r_- x)\, ,
\end{equation}
which still obey $f(x)=-g(-x)$. Here, $r_\pm = M \pm \sqrt{M^2-p^2}$, where $M$ is the mass and $|p|$ is the absolute value of the magnetic charges of the black holes. $r_{\pm}$ are the inner and outer black hole horizons (see below). $B$ is the magnitude of the external magnetic field. The range of variables is similar as before, with $\phi \in [-C\pi,C\pi]$ being periodic. The geometry (\ref{La8}) represents a pair of black holes of mass $M$ carrying opposite magnetic charges $\pm p$ immersed in an external magnetic field. Considering again coordinates (\ref{coor}), the geometry can be written as
\begin{equation}\label{C-metric-RN-Melvin}
ds^2 = \frac{\Lambda^2(r,\theta)}{(1+\alpha r \cos \theta )^2} \left( -F(r) dt^2 + \frac{dr^2}{F(r)} + \frac{r^2 d\theta^2}{G(\theta)} \right) + \frac{G(\theta) r^2 \sin^2 \theta }{\Lambda^2(r,\theta) (1+\alpha r \cos \theta )^2} d\phi^2 
\end{equation}
with
\begin{equation}
\Lambda(r,\theta) = \left( 1 - \frac{Bp}{2} \cos\theta \right)^2 + \frac{B^2r^2 \sin^2\theta }{4(1+\alpha r \cos \theta )^2} \hspace{0.3em} G(\theta),
\end{equation}
and
\begin{equation}
F(r) = (1 - \alpha^2 r^2) \Big{(}1- \frac{r_+}{r}\Big{)}\Big{(}1- \frac{r_-}{r}\Big{)},\quad \quad G(\theta) = (1 + \alpha r_+ \cos \theta )(1 + \alpha r_- \cos \theta ),
\end{equation}
where one sees that for $p=M=0$ the Melvin solution to Einstein-Maxwell equations is recovered, cf. \cite{Melvin}. When $B=0$, the solution becomes the magnetically charged analog of the $C$-metric (\ref{C-metric-RN}), and it reduces to the Reissner-Nordstr\"om solution when $\alpha = 0$. The case $\alpha=0\neq M$, on the other hand, is a particular case of the Ernst-Wild solution \cite{Ernst}, describing a black hole immersed in the Melvin universe. The gauge field configuration is given by
\begin{equation}
    A = - \frac{2}{B\Lambda(r,\theta)} \Big(1 - \frac{1}{2} B p \cos\theta \Big) \hspace{0.4em} d\phi
\end{equation}
which tends to the Dirac monopole configuration $A=(\text{const}+ p\cos \theta \, ) d\phi$ for small $B$.

By looking at the geometry near the poles $\theta =0$ and $\theta =\pi $, one finds that the values of $C$ that make the geometry regular are, respectively,
\begin{equation}
    C_{\theta = 0} = \frac{\Lambda^2(\theta=0)}{G(\theta=0)} \quad \text{and} \quad C_{\theta = \pi} = \frac{\Lambda^2(\theta=\pi)}{G(\theta=\pi)}.
\end{equation}
This means that, in order to cure the conical singularities at both poles, one has to fix the parameter $\alpha$ in terms of the product $pB$ as follows
\begin{equation}\label{alfa}
\Big({1-\frac 12 {pB}}\Big)^4   (1 - \alpha r_+)(1 - \alpha r_-)=\Big( {1+\frac 12 {pB}} \Big)^4 {(1 + \alpha r_+)(1 + \alpha r_-)}
\end{equation}
which in particular implies $C_{\theta = 0} = C_{\theta = \pi}$. The latter condition was impossible in the absence of the external magnetic field, but it is possible now provided $B$ takes the appropriate value. The physical interpretation of this is that, when (\ref{alfa}) is satisfied, the proper acceleration of the black holes is precisely that of a magnetic monopole with charge $p$ in an external magnetic field $B$, with no extra force needed. This explains why the cosmic string exactly cancels. Notice that, in the limit of small $\alpha $ condition (\ref{alfa}) yields the correct Newtonian limit
\begin{equation}
M\alpha\simeq pB,
\end{equation}
where we have used that $r_-r_+=p^2$ and $r_++r_-=2M$. Notice also that, as expected, condition (\ref{alfa}) is invariant under the transformation $pB, \alpha\to -pB,-\alpha $. 

From the solution above, having imposed the smoothness condition (\ref{alfa}), one can compute the production rate of black hole pairs in a magnetic field background. As said, this is the self-gravitating version of the magnetic Schwinger effect, for which the leading contribution to the creation rate is given by the standard instanton calculation
\begin{equation}\label{rate}
\Gamma \simeq e^{-2(I_{C}-I_{M})}
\end{equation}
where $I_C$ is the value of the on-shell Einstein-Hilbert action --with the appropriate Gibbons-Hawking-York boundary term-- evaluated on the $C$-metric solution (\ref{C-metric-RN-Melvin})-(\ref{alfa}) and regularized with respect to the action evaluated on the background geometry, $I_M$, the latter geometry being the Melvin solution \cite{Melvin} describing a uniform magnetic field $B$ in an otherwise empty spacetime. In the small charge limit, (\ref{rate}) reduces to the leading contribution of the standard Schwinger effect; see \cite{Grifits, Grifits2, Astorino}. In addition, it yields next-to-leading contributions that give an area-law term $\mathcal{A}/4$ in the series expansion of $\log \Gamma $, cf. \cite{Grifits2, Hawking}.

Another interesting case occurs when there is no acceleration; that is, the case in which the magnetically charged black hole is kept at fixed position despite the presence of the external magnetic field $B$. The presence of the cosmic string is unavoidable in that case, as it is precisely the string what makes the static configuration possible. Such a situation is described by the metric 
\begin{equation}\label{melvinR-Nmag}
    ds^{2} = \Lambda^{2}(r, \theta ) \left[ - F(r) dt^{2} + \frac{dr^{2}}{F(r)} + r^{2} d\theta^{2} \right] + \frac{r^{2} \sin^{2} \theta }{\Lambda^{2}(r, \theta )} d\phi^{2}
\end{equation}
with
\begin{equation}
    \Lambda (r,\theta )= 1 + \frac{B^{2}}{4} \left(r^{2} \sin^{2}\theta + p^{2} \cos^{2}\theta \, \right) - Bp \hspace{0.1em} \cos\theta  \ \ \text{and}  \ \ F(r) = 1 - \frac{2M}{r} + \frac{p^{2}}{r^{2}}\, .
\end{equation}
As before, $p$ is the magnetic charge of the monopole, $B$ is the external magnetic field, and $M$ is the mass; acceleration is zero in this case ($\alpha = 0$). The gauge field configuration is
\begin{equation}
    A = \frac{ \frac{B}{2}(r^2 \sin^2 \theta  + p^2 \cos^2 \theta \, ) -p \hspace{0.1em} \cos\theta }{\Lambda (r,\theta ) }  \hspace{0.2em} d\phi \, ,
\end{equation}
from what we observe that the particular case $B=0$ corresponds to the magnetically charged Reissner-Nordstr\"om solution. As in the previous cases, the parameter $C$ enters in the periodicity of the angular coordinate: $\phi \in[-C\pi,C\pi]$. The conditions for the absence of conical singularity at $\theta = 0$ and $\theta =\pi $ are now $C_{\theta = 0}=\Lambda^{2}(0)$ and $C_{\theta = \pi }=\Lambda^{2}(\pi )$, respectively; and we see that, provided $B\neq 0$, the condition $C_{\theta = 0}=C_{\theta = \pi }$ cannot be satisfied simultaneously. This is understood as the unavoidable presence of the cosmic string in the poles for the black hole to remain static. Notice that the pole at which the string is attached gets inverted if we perform the change $pB\to -pB$, as it can be easily seen from the explicit form of $\Lambda (r,\theta )$.

Solution (\ref{melvinR-Nmag}) can be thought of as a static, magnetically charged black hole in the Melvin universe \cite{Ernst}. Here, however, we are interested in solutions with non-vanishing acceleration, and so we will leave the study of stationary solutions of the Melvin type for a future work. So, let us go back to the accelerated black hole case: let us consider the electromagnetic dual to the configurations considered above. That is, consider the electrically charged black hole with constant proper acceleration. The solution takes a form quite similar to the $C$-metric; namely
\begin{equation}\label{Electrica}
    ds^2= \frac{1}{\alpha^2 (x+y)^2} \left( -f(y)\, d\tau^2 + \frac{dy^2}{f(y)} + \frac{dx^2}{g(x)} + g(x)d\phi^2 \right)
\end{equation}
with $f(y) = -(1-y^2)(1-\alpha r_+ y)(1-\alpha r_- y)$ and $g(x)=-f(-x)$. Now, $r_\pm = M \pm \sqrt{M^2-q^2}$ with $q$ being the electric charge: the electric potential reads $A = q y\, d\tau$. A change of coordinate similar to the one considered above yields the following form for the metric
\begin{equation}\label{C-metric-RN}
    ds^2 = \frac{1}{(1+\alpha r \cos \theta\, )^2} \left( -F(r)\, dt^2 + \frac{dr^2}{F(r)} + \frac{r^2 \, d\theta^2}{G(\theta)} + G(\theta)r^2 \sin^2\theta\, d\phi^2 \right)
\end{equation}
where now
\begin{equation}
F(r) = (1 - \alpha^2 r^2) \Big{(}1- \frac{r_+}{r}\Big{)}\Big{(}1- \frac{r_-}{r}\Big{)}\ ,\ \ \ G(\theta) = (1 + \alpha r_+ \cos \theta\, )(1 + \alpha r_- \cos\theta \, )\, ;
\end{equation}
the electric potential reads
\begin{equation}\label{Electrica2}
    A = \frac{q}{r}\, dt\, .
\end{equation}
This solution represents an accelerated version of the Reissner-Nordstr\"om black hole with a Rindler apparent horizon at $r_a=1/\alpha $, an outer black hole event horizon at $r_+= M + \sqrt{M^2-q^2}$ and an inner Killing horizon at $r_-= M - \sqrt{M^2-q^2}\leq r_+ \leq r_a $. As mentioned in the introduction, this electric $C$-metric solution has recently been considered in the context of the infinite-dimensional symmetries that the asymptotically locally flat spacetimes exhibit at null infinity. In \cite{Strominger}, the authors studied the action of the local BMS transformations on the $C$-metric. Here, we are interested in performing the near horizon analog of that analysis. That is, we are going to study how the infinite-dimensional symmetries that emerge in the near horizon region act on both the black hole horizons and the acceleration horizon of the $C$-metric. In particular, it will lead us to obtain the explicit form of the supertranslated, superrotated $C$-metric in the near horizon approximation.

\section{The near horizon symmetries of the $C$-metric}

Let us start by briefly reviewing the near horizon symmetries as studied in \cite{DGGP1} and in references thereof; see for example \cite{DGGP2, Cosmocute, Puhm, Charles}. To study the $C$-metric, unlike other cases, we find convenient to consider the set of boundary conditions proposed in \cite{Cosmocute}. This amounts to consider the set of metrics that, close to the horizon, behaves as follows
\begin{eqnarray}
 g_{tt} &=& - \kappa^{2} \rho^{2} + \mathcal{O}(\rho^{4}), \quad g_{t \rho} = \mathcal{O}(\rho^{3})  \nonumber \\
 g_{\rho \rho} &=& 1 + \mathcal{O}(\rho^{2}), \quad g_{AB} = \Omega_{AB}(\phi^{C}) + \mathcal{O}(\rho^{2}) \label{cosmologicalhorizons} \\
 g_{\rho A} &=& \mathcal{O} (\rho^{3}), \quad g_{tA} = \kappa N_{A}(\phi^{B}) \rho^{2} + \mathcal{O}(\rho^{4})
 \nonumber
\end{eqnarray}
where $\phi^{A}$ represent the angular variables, $A,B = 1,2$ (say $\phi^1 = \phi$, $\phi^2 = \theta $). $\rho \in \mathbb{R}_{\geq 0}$ is a radial coordinate that measures the separation from the horizon, the latter being located at $\rho =0$. The value of $\kappa$ gives the surface gravity at the horizon, and the functions $N_{A}(\phi^{B})$ and $\Omega_{AB}(\phi^{C})$ are arbitrary functions of the angular variables $\phi^1$ and $\phi^2$. $\mathcal{O}(\rho^n)$ stand for functions of the angles $\phi^A$ that damp off as fast as $\rho^n$, or faster, as $\rho $ tends to zero.

The set of metrics (\ref{cosmologicalhorizons}) is preserved by Killing vectors that respect the asymptotic conditions
\begin{align}\label{killing1}
& \mathcal{L}_{\xi} \hspace{0.2em} g_{tt} = \mathcal{O}(\rho^{4}) \quad , \quad \mathcal{L}_{\xi} \hspace{0.2em} g_{\rho \rho} = \mathcal{O}(\rho^{2}) \quad , \quad  \mathcal{L}_{\xi} \hspace{0.2em} g_{AB} = \delta \Omega_{AB}(\phi^{C}) + \mathcal{O} (\rho^{2})\\
 \nonumber
& \mathcal{L}_{\xi} \hspace{0.2em} g_{\rho A} = \mathcal{O}(\rho^{3}) \quad , \quad \mathcal{L}_{\xi} \hspace{0.2em} g_{\rho t} = \mathcal{O}(\rho^{3}) \quad ,  \quad \mathcal{L}_{\xi} \hspace{0.2em} g_{tA} = \kappa \, \delta N_{A}(\phi^{B}) \, \rho^{2} + \mathcal{O}(\rho^{4})
\end{align}
with $\delta \Omega_{AB}$ and $\delta N_{A}$ being arbitrary functions of $\phi^A$; here $\mathcal{L}_{\xi }$ stands for the Lie derivative with respect to the vector $\xi $. The asymptotic Killing vectors $\xi = \xi^{\mu}\partial_{\mu}$ obeying (\ref{killing1}) are those given by
\begin{align}\label{killing2}
    &\xi^{t} = T(\phi^{A}) + \mathcal{O}(\rho^{4}) \nonumber \\
    &\xi^{\rho} = \mathcal{O}(\rho^{4}) \\
    \nonumber
    &\xi^{A} = Y^{A}(\phi^{B}) + \mathcal{O}(\rho^{4})
\end{align}
where $T(\phi^{A})$, $Y^{1}(\phi^{A})$ and $Y^{2}(\phi^{A})$ are three arbitrary functions of the angles. These vectors form an infinite-dimensional algebra, which is the asymptotic isometry algebra at the horizon. The transformations defined by functions $T (\phi^{A})$ yield an Abelian ideal of the algebra; these are the so-called supertranslations, as they come to generalize the rigid $v$-translations $\xi=\partial_v$, the latter corresponding to $T=\text{const}$. The other transformations, the ones defined by the functions $Y^{A}(\phi^{B})$, form a non-Abelian subalgebra that can be taken to be the Witt algebra, and so they are interpreted as superrotations. In order to realize the full diffeomorphism algebra, it is convenient to choose complex coordinates $Z$, $\bar{Z}$ on the constant-$v$ sections of the horizon. These complex coordinates replace the angular variables $\phi^A$. Expanding functions $T$, $Y^{Z}$ and $Y^{\bar{Z}}$ in powers of $Z$ and $\bar{Z}$, one gets the vector basis
\begin{equation}\label{La27}
    \mathcal{Y}_{m} = Y_{m}^{Z} Z^{m+1} \partial_{Z} \quad , \quad \bar{\mathcal{Y}}_{m} = {Y}_{m}^{\bar{Z}} \bar{Z}^{m+1} \partial_{\bar{Z}}
\end{equation}
and
\begin{equation}\label{La28}
\mathcal{T}_{mn} = T_{(m,n)} Z^{m+1} \bar{Z}^{n+1} \partial_{t}\, ,
\end{equation}
where $Y_{m}^{A}$ ($A=Z,\bar{Z}$) and $T_{(m,n)}$ are Fourier coefficients. This yields the infinite-dimensional Lie algebra
\begin{align}\label{Algebrita}
    \left[\mathcal{Y}_{m},\mathcal{Y}_{n}\right] = (n - m) \hspace{0.2em}  \mathcal{Y}_{m+n} \quad , \quad &\left[\bar{\mathcal{Y}}_{m},\bar{\mathcal{Y}}_{n}\right] = (n - m) \hspace{0.2em} \bar{\mathcal{Y}}_{m+n} \quad , \quad \left[\mathcal{Y}_{m},\bar{\mathcal{Y}}_{n}\right] = 0 \\
    \nonumber
    \left[\mathcal{Y}_{p},\mathcal{T}_{mn}\right] = m \hspace{0.2em}  \mathcal{T}_{m+p\, n} \quad , \quad &\left[\bar{\mathcal{Y}}_{p},\mathcal{T}_{mn}\right] = n \hspace{0.2em} \mathcal{T}_{m\, n+p} \quad , \quad \left[\mathcal{T}_{pq},\mathcal{T}_{mn}\right] = 0 \, ,
\end{align}
with $m,n,p,q\in \mathbb{Z}$. This algebra consists of the semi-direct sum of two commuting copies of Witt algebra and an Abelian current algebra.

The Noether charges \cite{Glenn} associated to these symmetries are given by 
\begin{equation}\label{cargas2}
    Q \hspace{0.1em} [T,Y^A] = \frac{1}{16\pi } \int \, d^2\phi\, \sqrt{\det(\Omega_{AB})}\,  (2 \hspace{0.1em} \kappa \hspace{0.1em} T - Y^A N_A)\,   ,
\end{equation}
that can be decomposed in the charges $Q[T_{(m,n)},0]$, $Q[0,Y_m^A]$, the latter satisfying the same algebra as in (\ref{Algebrita}).

Now, let us show explicitly that the $C$-metric discussed above can be accommodated in the set of asymptotic horizon conditions (\ref{cosmologicalhorizons}). This would imply that such solutions also exhibit the infinite-dimensional asymptotic symmetry generated by (\ref{Algebrita}). In fact, by defining the radial coordinate
\begin{equation}\label{La31}
    \rho^2=\frac{4(r-r_0) \Lambda^2(r_0,\theta)}{F'(r_0)(1+\alpha r_0 \hspace{0.2em} \cos\theta \, )^2} ,
\end{equation}
where $r_0$ represents the location of a given horizon, we obtain
\begin{equation}
d\rho = \frac{1}{\sqrt{F'(r_0)}} \Big[ \frac{\Lambda (r_0,\theta)}{(1+\alpha r_0 \hspace{0.2em} \cos\theta\, )} \frac{dr}{\sqrt{r-r_0}} + 2 \sqrt{r-r_0} \hspace{0.4em} \frac{\partial}{\partial\theta}\Big(\frac{\Lambda (r_0,\theta)}{1+\alpha r_0 \hspace{0.2em} \cos\theta\, }\Big) \, d\theta \Big],
\end{equation}
and, with it, metric (\ref{C-metric-RN-Melvin}) takes the form
\begin{eqnarray}
    ds^2 &=& \Big[-\frac{F'(r_0)^2}{4} \rho^2 + \mathcal{O}(\rho^4) \Big] \, dt^2 + \Big[ 1 + \mathcal{O}(\rho^2) \Big]\, d\rho ^2 + \mathcal{O}(\rho^3)\, d\rho \, d\theta +  \\
\nonumber
 &&\Big[ \frac{r_0^2G^{-1}(\theta )\, \Lambda^2(r_0,\theta )}{(1+\alpha r_0\cos \theta )^2} + \mathcal{O}(\rho^2) \Big] \, d\theta^2 + \Big[ \frac{r_0^2G(\theta )\, \sin^2\theta }{(1+\alpha r_0\cos \theta )^2 \Lambda^2(r_0,\theta )}  + \mathcal{O}(\rho^2) \Big]\, d\phi^2 \, ,
\end{eqnarray}
from which we identify the form (\ref{cosmologicalhorizons}) with the particular functions
\begin{equation}
    \kappa = \frac{1}{2}F'(r_0), \quad N_A = 0, \quad \Omega_{\theta \theta} = \frac{r_0^2G^{-1}(\theta )\, \Lambda^2(r_0,\theta )}{(1+\alpha r_0\cos \theta )^2}, \quad \Omega_{\phi \phi} = \frac{r_0^2G(\theta )\, \Lambda^{-2}(r_0,\theta )\sin^2\theta }{(1+\alpha r_0\cos \theta )^2}, \quad \Omega_{\phi \theta} = 0\, ,
\end{equation}
with $A=\phi ,\theta $. Then, the near-horizon charges associated to the $C$-metric can be computed by evaluating (\ref{cargas2}). While the angular contributions vanish, namely $Q[0,Y^A] = 0$, the zero-mode of the supertranslation charge yields
\begin{equation}\label{TS_C}
    Q[1,0] = \frac{C}{4} \frac{F'(r_0) r_0^2}{1-\alpha^2 r_0^2} = \frac{\kappa }{2\pi} \frac{\mathcal A}{4} ,
\end{equation}
where we have used that the area of the constant-$t$, constant-$r$ section of the geometry (with $r=r_0$) is given by  
\begin{equation}\label{Arita}
    {\mathcal A} = \int_{-\pi C}^{\pi C} \int_0^{\pi} \sqrt{g_{\theta \theta} g_{\phi \phi}}_{|r=r_0} d\theta d\phi = \frac{4 \pi C r_0^2}{1-\alpha^2 r_0^2}.
\end{equation}

Charge (\ref{TS_C}) can actually be rewritten as $Q[1,0] =TS$, where $T = \kappa/2\pi$ is the Hawking temperature of the horizon located at $r_0 = r_a,r_{\pm}$ (here, $\hbar= k_B =1$), and $S={\mathcal A}/4$ is the Bekenstein-Hawking entropy associated to it. That is to say, the zero mode of the translation symmetry is actually computing the Wald entropy of the $C$-metric horizons. Evaluating (\ref{TS_C}) explicitly for the black hole Killing horizons $r_0=r_{\pm }$, we get
\begin{equation}\label{La37}
 Q[1,0]_{|r_{+}} =  - Q[1,0]_{|r_{-}} = \frac{C}{4} (r_+ - r_-) = \frac{(r_+ - r_-)}{4(1+\alpha r_+)(1+\alpha r_-)},
\end{equation}
which coincides with the result for the Reissner-Nordstr\"om black hole in the limit of small acceleration, $\alpha \to 0$. In (\ref{La37}), we have used that $\kappa $ for the horizon at $r_{+}$ is given by $\kappa = (r_+-r_-)(1-\alpha^2r_+^2)/(2r_+^2)$, and that $C$ can be taken to be $C_{\theta = 0}=G^{-1}(0)= (1+\alpha r_+)^{-1}(1+\alpha r_-)^{-1}$. Notice that the expressions for the charges above are valid for all the metrics considered here, provided one identifies the parameters accordingly. 

Next, we can consider the third horizon; namely, the apparent Rindler horizon located at $r_0 = r_a$. Of course, the result for the charge $Q[1,0]$ at $r_0 = 1/\alpha $ diverges, as it is easily seen from (\ref{Arita}); and this is easily understood as due to the fact that the Rindler wedge is non-compact. However, the charge per unit of area turns out to be finite and can be computed, yielding
\begin{equation}\label{Laultima}
 \vert Q[1,0]\vert _{|r_{a}}  /{\mathcal A} = \frac{\alpha }{8\pi }\, (1-\alpha {r_+}) \, (1-\alpha {r_-})\, .
\end{equation}
Notice that, in the limit $r_{\pm }\to 0$, this result tends to the correct result for the entropy per area of an acceleration horizon. In fact, (\ref{Laultima}) can be regarded as a self-gravitating generalization of the Laflamme result for the entropy per area for the Rindler horizon, cf. \cite{Laflam}, giving corrections to the Unruh temperature $T_{\text{U}}=\alpha /(2\pi ) + \mathcal{O}(\alpha^2 r_{\pm} )$. This also generalizes the near horizon computation of the Rindler density entropy done in \cite{DGGP2}. 

Apart from the charges associated to superrotations and supertranslations, in the case of the electrically charged $C$-metric, one can also consider a new set of infinite charges associated to gauge transformations that preserve the near horizon configuration \cite{Puhm}. By evaluating the zero modes of those charges for the solution (\ref{Electrica})-(\ref{Electrica2}), one finds that the electric charge as computed near the event horizon is given by   
\begin{equation}
Q_{U(1)} \, =\, \frac{q}{1+2 \alpha\, M+\alpha^2q^2}\, ,
\end{equation} 
which coincides with $q\, C$, $C$ being the periodicity of $\phi$ for the solution to be regular at $\theta = 0$.

\section{Superrotated, supertranslated $C$-metric}

In summary, we have shown here that the $C$-metrics can be accommodated in the near horizon boundary conditions (\ref{cosmologicalhorizons}) introduced in \cite{Cosmocute}. This means that the infinite set of symmetries generated by (\ref{Algebrita}) emerges in the vicinity of the horizons of such a solution. This is interesting as the $C$-metric exhibits features that were not present in the other configurations considered so far in similar contexts, the most salient one being the presence of the cosmic string that introduces a conical singularity at the horizon. Besides, the neutral and charged $C$-metric solutions exhibit isolated horizons of different types, so that having achieved a unified method to treat all such surfaces in an equal footing is interesting. We have seen that the horizon charges associated to the $C$-metric are non-trivial, with the zero-mode of the supertranslation correctly reproducing the Wald entropy of both Killing horizons and the Rindler horizon. 

Since we have proven that infinite-dimensional symmetries of the type considered in (\ref{killing2}) act on the $C$-metric preserving the near horizon structure (\ref{cosmologicalhorizons}), we can in principle compute the supertranslated and superrotated version of the $C$-metric. From (\ref{killing2}), one can see that, under an asymptotic diffeomorphism of the form (\ref{killing2}), the metric functions of the near horizon metric transform as follows 
\begin{equation}\label{La40}
   \delta \Omega_{AB} = \mathcal{L}_{Y}  \Omega_{AB} \quad , \quad \delta N_{A} = \mathcal{L}_Y N_A- \kappa \, \partial_{A} T\, ;
\end{equation}
therefore, by acting with Killing vector $\xi = h^A(\phi^B)\partial_A + f(\phi^B)\partial_v $ on the $C$-metric, we get the following geometry
\begin{eqnarray}
ds^2&=& \Big[ -\kappa^2 \rho^2 + \mathcal{O} (\rho^2)\Big]\, dt^2 + d\rho^2  - \Big[ 2 \kappa^2 \rho^2 \partial_A f + \mathcal{O}(\rho^4) \Big] \, d\phi^A dt +\nonumber \\
 &&\Big[\Omega_{AB} + h^C \partial_C \Omega_{AB} + \partial_Ah^C\, \Omega_{CB} + \partial_Bh^C\, \Omega_{AC} +\mathcal{O}(\rho^2)  \Big]  \, d\phi^A d\phi^B
\end{eqnarray}
with 
\begin{equation}
\Omega_{AB} = \frac{r_+^2  }{(1+\alpha r_+ \cos \theta )^2}
\Big[
\delta_{A\phi }\delta_{B \phi }\, {\Lambda^{-2}(r_0,\theta )\, G(\theta )\, \sin^2\theta } + 
              \delta_{A\theta }\delta_{B \theta }\, {\Lambda^2(r_0,\theta )\, G^{-1}(\theta )}
\Big]\, ,
\end{equation}
where $\phi^1=\phi$, $\phi^2=\theta $. Such is the near horizon form of the supertranslated, superrotated $C$-metric, which has superrotation and superrotation charges 
\begin{equation}\label{cargas22}
    Q \hspace{0.1em} [T,Y^A] = \frac{\kappa }{8\pi } \int \, d^2\phi\, \sqrt{\det(\hat{\Omega}_{AB})}\,  \Big( \, T f(\phi , \theta ) - \frac 12 Y^A\partial_{A} f(\phi , \theta )\, \Big)\,   ,
\end{equation}
with $\hat{\Omega}_{AB}= {\Omega}_{AB}+\mathcal{L}_h {\Omega}_{AB}$. Here, only the function $f(\phi , \theta ) $ appears as $h^A(\phi , \theta )$ enters in the charge only when the original metric, the one before applying the superrotation and supertranslation, has non-vanshing $N^A$. This is clear from (\ref{La40}). Still, a non-vanishing superrotation charge appears in (\ref{cargas22}) due to the supertranslation generated by $h(\phi , \theta )$. This expresses at the level of the charges the fact that supertranslations and superrotations do not commute in the charge algebra (\ref{Algebrita}). Notice that $h^A(\phi , \theta )$ and $f(\phi , \theta )$ represent here what $Y^A(\phi , \theta )$ and $T(\phi , \theta )$ were in (\ref{La40}). Here, it was necessary to introduce a new notation for not to mistake the functions $h^A(\phi , \theta )$, $f(\phi , \theta )$ that parameterize the superrotation and supertranslation for the functions $Y^A(\phi , \theta )$, $T(\phi , \theta )$ that define the Killing vector with respect to which the charge (\ref{cargas22}) is computed.

The set of suitable functions $f(\phi , \theta )$ is defined by taking into account the points where the horizon surface is smooth and the points where it presents singularities. While, in virtue of the generality of the results of \cite{DGGP1}, we expected the horizon of accelerated black holes to exhibit the infinite symmetries (\ref{La27})-(\ref{La28}) locally, globally the problem which we deal with herein is different. In the case of the $C$-metric, there exists a global obstruction due to the presence of a conical singularity at the point where the cosmic string pinches the horizon. This introduces two differences with respect to the case of smooth horizons. On the one hand, the functions $Y^A (\phi , \theta )$, $T(\phi , \theta )$ with respect to which the charges (\ref{cargas22}) are defined are in principle allowed to admit singularities at the point where the cosmic string is located. On the other hand, the functions $h^A (\phi , \theta )$, $f(\phi , \theta )$ considered to superrotate and supertranslate the original $C$-metric are now to be restricted to the set of functions that preserve the location of the conical singularity, $\theta = \pi $. This means that, even though local properties will be shared with the smooth horizons, as the fact that the algebras agree manifestly shows, the global transformations will be different. A direct way to see this is to look at the expression (\ref{La31}), which gives the change of coordinates that suffices to put the $C$-metric in its near horizon form. Such change of coordinates is in general well defined except at $\theta = \pi $, meaning that the analysis will be such that the transformations to be considered have to preserve the location of the conical singularity.  

Our result of the infinite-dimensional symmetries of the accelerated horizons acquires particular interest when considered in relation to the analysis done in \cite{Strominger}. In the latter work, the BMS symmetries of the $C$-metric at null infinity were studied, while here we have presented the horizon version of the story. This suggests to perform a comparative analysis similar to the one done in \cite{Puhm}: In \cite{Puhm}, the memory effect produced at a black hole horizon by a transient gravitational shockwave was computed. It had already been shown in \cite{HPS} that such a gravitational wave produces a deformation of the black hole geometry which from future null infinity is seen as a BMS supertranslation, and \cite{Puhm} gives the complementary description of such a physical process as seen from the horizon perspective. This showed that, for an observer hovering just outside the event horizon, in addition to a supertranslation the shockwave also induces a superrotation. Considering our results herein in combination with the analysis of \cite{Strominger} would make possible, in principle, to extend the comparison between the asymptotic region and the near horizon region to the case of $C$-metric type solutions. However, this is more involved than in the cases of static black holes considered in \cite{Puhm}. We plan to address this in future work. 

Now, with the intention of showing that the analysis performed here is general enough and can actually be extended to other cases of interest, we will dedicate the next sections to show how the local change of coordinates (\ref{La31}) can be applied to other solutions. First, we will consider the case of accelerated black holes in AdS space, which exhibits features that are qualitatively different \cite{Holo0, unotro1, Krtous:2005ej}. Second, we will consider a system describing binary black holes at equilibrium \cite{AstorinoBinary1}.

\section{Accelerated black holes in AdS}

Let us consider black holes accelerating in AdS spacetime \cite{Holo0}. This is given by the AdS $C$-metric
\begin{equation}\label{tango1}
    ds^2 = \frac{1}{\Omega^2(\theta )} \left( -F(r) \, dt^2 + \frac{dr^2}{F(r)} + \frac{r^2 }{G(\theta)}\, d\theta^2 + \frac{G(\theta)}{K^2}\, r^2 \sin^2\theta \, {d\phi^2} \right)
\end{equation}
with
\begin{equation}
    \Omega (\theta )= 1+ \alpha r \cos\theta , \quad K = 1 + 2m\alpha ,
\end{equation}
and
\begin{equation}    \label{tango3}
F(r) = (1 - \alpha^2 r^2) \Big{(}1-\frac{2m}{r}\Big{)} + \frac{r^2}{\ell^2}, \quad G(\theta) = 1 + 2m \alpha \cos\theta
\end{equation}
where $\ell = \sqrt{-3/\Lambda }$ is the radius of AdS, $\Lambda <0$ is the cosmological constant, $\alpha $ is the black hole acceleration, and $m $ is related to the black hole mass. When $\alpha = 0 $ solution (\ref{tango1})-(\ref{tango3}) reduces to the AdS-Schwarzschild black hole geometry. The value of $K$ is chosen for the metric to be regular everywhere except at the point $\theta = 0$.

In contrast to accelerated black holes in flat space or in de Sitter space, the case with negative cosmological constant exhibits peculiar features and a richer causal structure that depends on the value of the acceleration relative to the inverse of the AdS radius \cite{unotro1}. For $\alpha^2 > 1/\ell^2$ there is more than one horizon, corresponding to the black hole ($r_+$) and the acceleration horizon ($r_a$); in contrast, for $\alpha^2 < 1/\ell^2$ only the black hole horizon exists. In other words, while for acceleration values $\alpha^2 > 1/\ell^2$ the $C$-metric in AdS represents a pair of black holes accelerated in opposite directions, for $\alpha^2 < 1/\ell^2$ the solution describes a single accelerated black hole in AdS. That is, only if $\alpha $ is above certain threshold the solution in AdS space exhibits the features of an actual acceleration, and an effective acceleration $\sqrt{\alpha^2-\ell^{-2}}$ appears. This can be easily understood by interpreting AdS as a decelerating universe with a cosmic acceleration $\Lambda /3$. Another way of interpreting this is by noticing that the Unruh temperature of an accelerated observed in AdS only exists provided $\alpha > 1/\ell^2$, its expression being \cite{Deser, Jacobson}
\begin{equation}
T_{\text{U}}= \frac{1}{2\pi }\sqrt{\alpha^2-\ell^{-2}}\, .    
\end{equation}

The fact that the $C$-metric with $\alpha^2 < 1/\ell^2$ describes a single accelerated black hole makes it simpler to address this case. This is because in the case $\alpha^2 > 1/\ell^2$, due to the presence of the Rindler type horizon that separates the two black holes, the asymptotic region at infinity changes, and this requires a special care when computing conserved charges resorting to standard methods. The conserved charges in the case $\alpha^2 < 1/\ell^2$ were computed in \cite{Holo1}. It was observed there that, due to the non-vanishing acceleration of the frame in which the black hole is at equilibrium with respect to the AdS boundary, it is necessary to correctly identify the normalization of the asymptotic timelike Killing vector with respect to which the mass is computed. The result thus differs from the {\it naive} calculation by a factor $\sqrt{1-\alpha^2\ell^2}$; see \cite{Holo1} for details. Here, we can extend that calculation and address the case $\alpha^2 > 1/\ell^2$ in a unified framework. Our method enables to do that because it relies on the near horizon region rather than in the near boundary asymptotic.

We considering the change of coordinates
\begin{equation}
    \rho^2=\frac{4(r-r_0)}{F'(r_0)(1+ \alpha r_0 \hspace{0.2em} \cos \theta \, )^2} ,
\end{equation}
which leads to the near horizon form
\begin{eqnarray}
    ds^2 &=& \Big[-\frac{F'(r_0)^2}{4} \rho^2 + \mathcal{O}(\rho^4) \Big] \, dt^2 + \Big[ 1 + \mathcal{O}(\rho^2) \Big]\, d\rho^2 + \mathcal{O}(\rho^3)\, d\rho\, d\theta +  \\
\nonumber
 &&\Big[ \frac{r_0^2\, G^{-1}(\theta )}{(1+ \alpha r_0\cos \theta )^2} + \mathcal{O}(\rho^2) \Big] \, d\theta^2 + K^{-2}\Big[ \frac{r_0^2 \, G(\theta )\sin^2\theta }{(1+ \alpha r_0\cos \theta )^2}  + \mathcal{O}(\rho^2) \Big]\, {d\phi^2} \, .
\end{eqnarray}
This yields the metric functions
\begin{align}\label{funciones}
    &\kappa = \frac{1}{2}F'(r_0), \ \ \quad N_{\theta} =0,\ \ \quad N_{\phi} = 0,\ \ \quad \Omega_{\phi \theta} = 0\\
    \nonumber
    &\Omega_{\theta \theta} = \frac{r_0^2\, G^{-1}(\theta )}{(1+ \alpha r_0\cos \theta )^2}, \quad \Omega_{\phi \phi} = \frac{r_0^2\, G(\theta )\sin^2\theta }{K^2 (1+\alpha r_0\cos \theta )^2} \, .
\end{align}

Next, considering asymptotic Killing vectors $\xi = \xi^{\mu}\partial_{\mu}$ that preserve the near horizon form, namely with $\xi^{t} = T(\phi^{A}) + \mathcal{O}(\rho^{4})$, $\xi^{\rho} = \mathcal{O}(\rho^{4}) $, $\xi^{A} = Y^{A}(\phi^{B}) + \mathcal{O}(\rho^{4})$, and evaluating the charges (\ref{cargas2}), we find
\begin{equation}\label{la82}
    Q[T=t_0,0] = \frac{t_0}{4 K} \frac{F'(r_0) r_0^2}{(1-A^2 r_0^2)} =  \frac{t_0 \kappa }{2\pi} \frac{\mathcal A}{4} = TS
\end{equation}
where we have taken $Y^A=0, T =t_0$, with $t_0$ being a constant that controls the normalization of the null Killing vector $\sim \partial_t$ at the horizon. We also used that  
\begin{equation}
    F'(r_0) = \frac{2m}{r_0^2} + 2 \alpha^2 (m - r_0) + \frac{2 r_0}{\ell^2}.
\end{equation}
For the choice $t_0=1$ one obtains the result 
\begin{equation}
     Q[1,0] =  TS = \frac{m}{2(1+2\alpha m)} + \frac{r_0^3}{2\ell^2 (1+2\alpha m) (1- \alpha^2 r_0^2)^2}
\end{equation}
which, when evaluated on the black hole event horizon $r_0=r_+$, exactly agrees with the result for the product of the entropy and the temperature obtained in \cite{C2}. However, as mentioned above, it was noticed in \cite{Holo1} that in the case $\alpha^2<1/\ell^2$ a more convenient normalization is $t_0=\sqrt{1-\alpha^2\ell^2}$. With this choice, our result agrees with that of \cite{Holo1}. It is worth emphasizing that the computation we presented here, being done near the horizon and so dispense with the near boundary region, is valid also for the range $\alpha^2>1/\ell^2$ provided one normalizes as $t_0=1$, i.e. computing the charge with respect to the Killing vector $\partial_t$ at the horizon.

\section{Black hole binary system}

Another case that can be explicitly solved with our near horizon analysis is the solution recently found in \cite{AstorinoBinary1}, which describes a binary black hole system at equilibrium. This is an exact solution to 4-dimensional Einstein equations in vacuum that represents a symmetric or asymmetric pair of static black holes at equilibrium. The metric is completely regular at the event horizons, implying that the balance between the two Schwarzschild-like sources is achieved by the presence of an external gravitational field, without additional external fields, nor strings or struts. This can be regarded as a purely gravitational analog of the magnetically charged Ernst-Wild black hole in the Melvin universe, cf. \cite{Brenner}.

Although analytically tractable, the metric of \cite{AstorinoBinary1} takes a cumbersome form, so we will not write it down here; instead, we refer the reader to the original paper; see also \cite{AstorinoBinary2}. For us, it will be sufficient to consider the form the metric takes near the horizon of one of the two black holes; that is
\begin{equation}
    ds^2 \simeq h(\theta) \left( -f(r) e^{F_1(\theta)}\, dt^2 + D^2 e^{F_2(\theta)}\, \frac{dr^2}{f(r)} + r_0^2 D^2 h(\theta) e^{F_2(\theta)}  \, d\theta^2 + r_0^2 e^{-F_1(\theta)}  \frac{\sin^2\theta}{h(\theta)}\,  d\phi^2 \right)
\end{equation}
with
\begin{equation}
    h(\theta) = \frac{m_1 \cos\theta + m_2 + z_1 - z_2}{m_1 \cos\theta - m_2 + z_1 - z_2}, \ \ \ \ f(r) = 1 - \frac{2m_1}{r}, \ \ \ \ D = \frac{m_1 + m_2 - z_1 + z_2}{m_1 - m_2 - z_1 + z_2},\nonumber
\end{equation}
and
\begin{eqnarray}
    F_1(\theta) &=& 2 (b_1 + b_2 z_2 + b_2 m_1 \cos \theta ) (z_1 + m_1 \cos \theta )\nonumber \\
    \nonumber
    F_2(\theta) &=& 2b_1 (m_1 \cos \theta - 2m_1 - 4m_2 - z_1) + 2b_2(m_1^2 \cos^2 \theta \nonumber\\
    &&+ 2m_1 z_1 \cos \theta - 2m_1^2 - z_1^2 - 4m_1 z_1 - 8m_2 z_2),\nonumber
\end{eqnarray}
$b_1, \, b_2, \, m_1, \, m_2, \, z_1, \, z_2$ being parameters of the solution, whose physical meaning are discussed in \cite{AstorinoBinary2}. In particular, the location of the horizon of one of the two black holes described by the solution above is given by $r_0 = 2 m_1$. Considering the change of coordinates
\begin{equation}
    \rho^2=\frac{4(r-r_0)}{f'(r_0)} h(\theta) D^2 e^{F_2(\theta)} ,
\end{equation}
for $r_0=2m_1$, so that $\rho = 0$ at the event horizon of one of the black holes, we arrive to the near horizon form
\begin{eqnarray}
    ds^2 &=& \Big[-\frac{f'(r_0)^2}{4} \frac{e^{F_1(\theta)-F_2(\theta)}}{D^2} \rho^2 + \mathcal{O}(\rho^4) \Big] \, dt^2 + \Big[ 1 + \mathcal{O}(\rho^2) \Big]\, d\rho^2 + \mathcal{O}(\rho^3)\, d\rho\, d\theta +  \\
\nonumber
 &&\Big[ r_0^2 D^2 h(\theta) e^{F_2(\theta)} + \mathcal{O}(\rho^2) \Big] \, d\theta^2 + \Big[ r_0^2 e^{-F_1(\theta)}  \frac{\sin^2 \theta }{h(\theta)}  + \mathcal{O}(\rho^2) \Big]\, \frac{d\phi^2}{K^2} \, ,
\end{eqnarray}
which actually satisfies the near horizon boundary conditions discussed above, with the metric functions being
\begin{align}\label{funciones}
    &\kappa =  \frac{f'(r_0)}{2D}e^{\frac{F_1(\theta)-F_2(\theta)}{2}}, \quad N_{\theta} =0, \quad N_{\phi} = 0, \quad \Omega_{\phi \theta} = 0\, , \nonumber \\
    \nonumber
    &\ \ \ \ \Omega_{\theta \theta} =  r_0^2 D^2 h(\theta) e^{F_2(\theta)}, \quad \Omega_{\phi \phi} = r_0^2 e^{-F_1(\theta)}  \frac{\sin^2 \theta }{h(\theta)}\, .
\end{align}

The fact of having achieved to put the metric of \cite{AstorinoBinary1} in the convenient boundary conditions to perform the near horizon analysis is remarkable, as that solution takes a very complicated form. Having achieved so enables us to work out the thermodynamics of the solution from a near horizon perspective. From the surface gravity $\kappa $ obtained above we can get the temperature --up to a normalization $t_0$ of the null Killing vector at the horizon--, yielding
\begin{equation}
    T_1 = \frac{\kappa}{2\pi} = \frac{t_0}{8 \pi m_1}  \frac{(m_1 - m_2 - z_1 + z_2)}{(m_1 + m_2 - z_1 + z_2)} \, e^{2 b_1(2m_2 + m_1 + z_1) + 2b_2((m_1+z_1)^2 + 4m_2 z_2)}\, ,
\end{equation}
which agrees with the analysis in \cite{AstorinoBinary1} --there, $t_0$ is denoted $\alpha$--. By considering the asymptotic Killing vectors that preserve the form at the horizon, we can compute the charge for a black hole of the pair, and we obtain\begin{equation}\label{la82}
    Q[T=t_0,Y^A=0] = \frac{t_0 \kappa }{2\pi} \frac{\mathcal A}{4}  
\end{equation}
which, consistently, yields
\begin{equation}
  Q[T=t_0,Y^A=0] =  T_1 S_1 = \frac{t_0 m_1}{2}\, .
\end{equation}
Considering the mass of the black hole to be $M_1 = t_0 m_1$, this is found to satisfy the Smarr formula $\frac{1}{2}M_1 = T_1 S_1$. Therefore, our near horizon analysis is completely consistent with the results of \cite{AstorinoBinary1, AstorinoBinary2} and proves to be powerful enough to be applied to physical scenarios described by quite different solutions. Other solutions can also be analyzed in this framework, like the one considered in \cite{Astor}; see also \cite{Brenner}. In particular, we expect to come back soon to the problem of analyzing the family  of solutions studied in \cite{AstorinoBinary1, AstorinoBinary2, AstorinoBinary3}, which comprises many different cases, including binary systems of rotating black holes.

\[\]

The authors thank Marco Astorino, Laura Donnay, Hern\'{a}n Gonz\'{a}lez, Julio Oliva, Miguel Pino, Andrea Puhm, and Adriano Vigan\`{o} for discussions. This work has been supported by CONICET and ANPCyT through grants PIP-1109-2017, PICT-2019-00303. The research of AA is supported in part by Proyecto de cooperaci\'{o}n internacional 2019/13231-7 FAPESP/ANID and by the Fondecyt Grants 1210635, 1181047 and 1200986.

  \end{document}